\documentclass{emulateapj}
\usepackage{epsfig}
\usepackage{apjfonts}
\usepackage{epsfig}
\usepackage{epsfig}

\begin{document}

\newcommand{\tpass}{{t$_{\rm pass}$}}
\newcommand{\DeltaD}{{$\Delta D$}}
\newcommand{\DeltaV}{{$\Delta V$}}
\newcommand{\Nhost}{{\rm N}}
\newcommand{\Nhalo}{{\rm N}}
\newcommand{\DN}{${\rm D_N}$}
\newcommand{\ewha}{${\rm EW(H\alpha)}$}
\newcommand{\Vin}{${\rm V_{in}}$}
\newcommand{\Vnow}{${\rm V_{now}}$}\
\newcommand{\Vmax}{${\rm V_{\rm max}}$}
\newcommand{\Nseven}{${\rm N_{700}}$}
\newcommand{\kms}{km s$^{-1}$}

\shorttitle{Isolating Triggered Star Formation}
\shortauthors{Barton et al.}

\bibliographystyle{apj}

\title{Isolating Triggered Star Formation}

\author{\sc Elizabeth J. Barton\altaffilmark{1}, Jacob A. Arnold\altaffilmark{1},
Andrew R. Zentner\altaffilmark{2,3,4}, James S. Bullock\altaffilmark{1}, and
Risa H. Wechsler\altaffilmark{5}}

\altaffiltext{1}{Center for Cosmology, Department of Physics and
Astronomy, University of California, Irvine,  CA 92697-4575
(email: ebarton@uci.edu)}

\altaffiltext{2}{Kavli Institute for Cosmological Physics, 
Department of Astronomy and Astrophysics, and The Enrico Fermi Institute, 
The University of Chicago, 
Chicago, IL 60637}

\altaffiltext{3}{Present Address: Department of Physics and Astronomy
University of Pittsburgh, Pittsburgh, PA 15260, zentner@pitt.edu}

\altaffiltext{4}{National Science Foundation Fellow}

\altaffiltext{5}{Kavli Institute for Particle and Astrophysics \& 
Cosmology, Physics Department, and Stanford Linear Accelerator Center,
Stanford University, Stanford, CA 94305}

\begin{abstract}

Galaxy pairs provide a potentially powerful means of studying
triggered star formation from galaxy interactions.  We use a large
cosmological N-body simulation coupled with a well-tested
semi-analytic substructure model to demonstrate that the majority of
galaxies in close pairs reside within cluster or group-size halos and
therefore represent a biased population, poorly suited for direct
comparison to ``field'' galaxies.  Thus, the frequent observation that
some types of galaxies in pairs have redder colors than ``field''
galaxies is primarily a selection effect.  We use our simulations to
devise a means to select galaxy pairs that are isolated in their dark
matter halos with respect to other massive subhalos (\Nhalo $=2$
halos) and to select a control sample of isolated galaxies (\Nhalo
$=1$ halos) for comparison.  We then apply these selection criteria to
a volume-limited subset of the 2dF Galaxy Redshift Survey with M$_{\rm
B, j} \leq -19$ and obtain the first clean measure of the {\it
typical} fraction of galaxies affected by triggered star formation and
the average elevation in the star formation rate.  We find that 24\%
(30.5 \%) of these L$^\star$ and sub-L$^{\star}$ galaxies in isolated
50 (30) h$^{-1}$ kpc pairs exhibit star formation that is boosted by a
factor of $\gtrsim 5$ above their average past value, while only 10\%
of isolated galaxies in the control sample show this level of
enhancement.  Thus, 14\% (20 \%) of the galaxies in these close pairs
show clear triggered star formation.  Our orbit models suggest that
12\% (16\%) of 50 (30) h$^{-1}$ kpc close pairs that are isolated
according to our definition have had a close ($\leq$ 30 h$^{-1}$ kpc) pass
within the last Gyr.  Thus, the data are broadly consistent with a
scenario in which most or all close passes of isolated pairs result in
triggered star formation.  The isolation criteria we develop provide a
means to constrain star formation and feedback prescriptions in
hydrodynamic simulations and a very general method of understanding
the importance of triggered star formation in a cosmological context.

\end{abstract}


\section{Introduction} \label{sec:intro}

Galaxy interactions and mergers help drive galaxy evolution.  In the
concordance $\Lambda$CDM cosmological model, nearly every galaxy has
had a major merger at least once over cosmic time
\citep[e.g.,][]{Maller06,Stewart07}.  Major mergers and
interactions consume available gas, producing stellar populations 
which subsequently redden with age
\citep[e.g.,][]{Larson78, Kennicutt87}. Major mergers destroy disks,
turning galaxies into spheroids \citep[e.g.,][]{Toomre72},
but also potentially creating large disk galaxies in gas-rich 
scenarios \citep[e.g.,][]{Robertson06}.  Close
galaxy passes and minor mergers are even more frequent.  These
perturbations send gas into the centers of galaxies, where they may
contribute to bulge components and even feed black holes
\citep[e.g.,][]{Kennicutt84,Barnes92, Mihos96, Barton00, Combes01,
Barton01, Barton03, Kannappan04, Kormendy04,FreedmanWoods06,Lin07,
SolAlonso07, FreedmanWoods07}.  In the hierarchical picture of galaxy
formation, mergers play an even more important role during earlier epochs
than they do today, possibly triggering the high star formation rates
\citep{Lowenthal97, Kolatt99, Somerville01, Wechsler01} observed in 
some ``Lyman break'' galaxies at $z\gtrsim3$ \citep{Steidel92, Steidel96}
or luminous submillimeter galaxies \citep[e.g.,][]{Chapman05}.

Existing detailed studies of galaxy interactions show that triggered
star formation and morphological evolution can be very rapid and
intense.  However, we can only observe a ``snapshot'' of the evolution
of galaxy pairs.  With incomplete phase-space information, we do not
know their true frequency, evolutionary timescales, orbits, or fates.
Thus, we have neither a complete understanding of the frequency of
interactions and mergers nor knowledge of their impact on galaxy
evolution.  Many processes such as ram pressure stripping, interaction
with a cluster or group potential, or high-speed ``harassment,'' may
also be very important in establishing the morphologies and stellar
populations of galaxies, and in driving the relationships between
morphology, color, and local environmental density
\citep[e.g.,][]{Gunn72, Dressler80, Postman84, Moore96,Blanton05a}.

Hydrodynamic simulations of galaxy interactions can make predictions
for the expected star formation rate from any type of galactic
collision \citep[e.g.,][]{Barnes92,Mihos96,Springel05, Cox06, Perez06a,
Perez06b}.
However, these predictions are sensitive to prescriptions for baryonic 
physics that are highly uncertain.  In particular, hydrodynamic simulations 
must rely on effective models for gas dynamics, 
cooling, and star formation that these simulations 
cannot hope to resolve.  Uncertainties in these effective theories 
are an enormous problem for galaxy formation theory and detailed tests
of these prescriptions are required.  Unfortunately 
such tests are not simple; the full phase-space 
information is not known for many real systems.
However, empirical, {\it statistical} studies of galaxy interactions and
mergers --- when combined with predictions of galaxy orbits in a 
now well-established cosmological model --- hold promise for uncovering the role of 
interactions and mergers in galaxy evolution.

The first step in studying the effects of interactions and mergers is
to identify systems that are undergoing these processes.  Using
morphological distortion as an indicator of galaxy interactions 
does yield a subset of the pairs that have definitely had a
close pass.  However, {\it relying on tidal distortion leads to missed
interacting pairs} because (1) sensitivity to low-surface-brightness
features is a strong function of seeing, depth, and redshift, (2) the
morphological features are short-lived ($\sim$100 Myr), and (3) many
of the features, like tidal tails, are resonance effects that only
appear in prograde disk galaxy encounters \citep{Toomre72}.
For a more complete census of the effects of interactions, it is important 
to select pairs or systems of galaxies based {\it only} on
proximity in redshift and projected separation, then to use 
mock catalogs constructed from cosmological
simulations to understand this selection.

Historically, studies of distorted galaxies and galaxies in pairs have
been very revealing.  The early stages of galaxy interactions can
drive gas into the center of a progenitor galaxy and trigger an early
episode of central star formation long before the final merger
\citep[e.g.,][]{Larson78, Joseph84, Kennicutt87, Mihos94, Mihos96}.
The strength of the optical emission line associated with this star
formation correlates with the separation of the pair on the sky
(\DeltaD) and in redshift \citep[\DeltaV;][]{Barton00}.  The
optically-detected star formation is strongest in the central few kpc,
often dominating the optical light; in the case of a late-type spiral,
it can also contribute substantially to the formation of a bulge
\citep{Tissera02, Barton03}.  This process may be the primary
mechanism for the formation of late-type bulges \citep{Kannappan04}.
\citet{Kewley06} show that the metallicities of these galaxies provide
a ``smoking gun'' for gas infall from the outskirts of the disks.
Galaxies with (optically) strong central starbursts have metallicities
that are lower than average for their luminosities, consistent with a
starburst occurring in gas that was driven into the nucleus from the
metal poor outskirts of the progenitor's disk.

New, large redshift surveys such as the 2dFGRS \citep{Colless01} and
Sloan Digital Sky Survey \citep{York00} provide a means of exploring 
ever-larger samples of galaxies in pairs, although these samples must be
approached with caution because mechanical spectrograph constraints
make them deficient in close pairs. The correlations between orbital
parameters and star-forming properties of galaxies in pairs have been
verified in both the 2dFGRS \citep{Lambas03}, and the SDSS
\citep{Nikolic04, Luo07}.  However, the studies also appear to reveal
unexpected phenomena including apparently {\it suppressed} star
formation in widely-separated pairs and red tails in the color
distributions of paired galaxies \citep{Lambas03, SolAlonso04,
SolAlonso06,Luo07}.

Using these empirical results to arrive at a true measure of the
amount and timescales of triggered star formation is difficult.  Pair
samples are complex: they include some interlopers and are rich in
galaxies in a variety of environments \citep[e.g.,][]{SolAlonso06,
Soares07}. The progenitors of galaxy interactions come from a mixture
of galaxy types \citep[e.g.,][]{Focardi06}.  In dense environments,
many will have already experienced multiple close passes and mergers.
They may have consumed or lost much of their gas, leaving little to
form new stars.  However, in sparser environments, the progenitors of
the interaction may have had very few previous interactions.  They may
have large remaining gas reservoirs.  Thus, the efficacy of an
interaction in triggering star formation should depend in detail on
the environment of the interacting galaxies.

Dense clusters of galaxies are straightforward to identify with
confidence.  However, the low-speed encounters that trigger star
formation are extremely unlikely in these clusters.  An understanding
of tidally-triggered star formation requires accurate probes of
isolated galaxies and sparse loose groups.  Unfortunately, accurate
environmental statistics are difficult to interpret precisely in
this regime.  Group-finding algorithms are effective at investigating
environments \citep{Huchra82}.  However, even highly tuned group-finding
algorithms construct false groups and miss group members
\citep[e.g.,][]{Yang05, Gerke05, Weinmann06, Berlind06, Koester07}.
Nearest-neighbor and counts-in-cylinder statistics also have a large
scatter \citep[e.g.,][]{Berrier07}.

The use of mock galaxy catalogs based on numerical simulations of
structure formation in the standard cosmology can help sort these
issues out.  Direct numerical simulation is difficult because
obtaining the necessary resolution to be complete in close pairs while
simultaneously modeling a large enough volume to reduce sample
variance is computationally expensive.  An alternative and proven
method is to use an analytic model for dark matter halo substructure
\citep[so-called ``subhalos'', e.g.,][]{Zentner03,Taylor05,Zentner05}
in conjunction with an $N$-body simulation of a
cosmologically-relevant volume.  The analytic model treats halo
substructure using a method that is approximate, but free of inherent
resolution limits, and extends the effective numerical resolution in
the simulated volume, ensuring completeness in the dense environments
where many close pairs reside.  The analytic model also allows for a
quantification of shot-noise contributions to close-pair samples that
may arise, for example, from the particulars of galaxy orbits and may
be significant \citep{Berrier06}.  Such models have been developed and
validated by comparison to direct $N$-body simulations in the regimes
where the two techniques are commensurable
\citep{Zentner05}.  Thus, the time is
ripe to develop new methods aimed at understanding the detailed
make-up of pairs of galaxies selected from redshift surveys.

We study mock catalogs constructed using the hybrid approach 
described briefly in the previous paragraph and extensively in 
\citet{Zentner05} and \citet{Berrier06}. This approach has
already succeeded in explaining the apparent lack of evolution in the
close pair fraction observed in intermediate-redshift surveys
\citep{Lin04, Berrier06}.  In the present study, we focus on interpreting 
the star-forming properties of galaxies in pairs.  With the
complexities of the large-scale environments of pairs in mind, we
explore the construction of appropriate samples and control samples for galaxies in
pairs.  In particular, we examine ways to isolate the {\it immediate} effects
of triggered star formation from other environmental processes.  Here,
we apply the analysis to a volume-limited sample of galaxies in the
2dFGRS; however, the techniques we discuss are generally applicable to
other studies of tight sub-groupings of galaxies.

In \S~\ref{sec:methods} we describe the numerical methods and the 
observational data that we use.  
\S~\ref{sec:environments} contains a description of the
model predictions for the environments of galaxies in pairs; in this
section we examine the problem of trying to construct a control sample
of objects that are not in pairs.  In \S~\ref{sec:isolated}, we
restrict to the most isolated pairs and individual galaxies and show
that isolated pairs are almost purely dark matter halos containing
\Nhalo\ $=2$ galaxies inside their virial radii. The isolated
galaxies we select are an appropriate control sample for the progenitors
of an interaction.  We examine galaxies in pairs in the 2dF Galaxy
Redshift Survey \citep[2dFGRS][]{Colless01} in \S~\ref{sec:data} and
measure the amount of triggered star formation in isolated pairs
relative to the control.  We \S~\ref{sec:complex} contains a brief
description of the situation in more complex environments, and we
conclude in \S~\ref{sec:conclusion}.  In future papers, we plan to
explore more complex environments in greater detail.  The ultimate
goal of this and similar studies is to isolate the effects of
interactions and then re-integrate the results into a complete
cosmological interaction history for galaxies, thus measuring the
amount of galaxy evolution triggered by interactions and mergers.

\section{The Models and the Data} \label{sec:methods}

The first step in our study is to construct mock catalogs of galaxies and
study the environments of galaxies in projected, close-pair
configurations in order to understand the selection of such objects.
We use the information gleaned from the
mock samples to aid in the interpretation of close pairs in the
2dFGRS.  In this section, we describe the models used to produce mock
catalogs and the observational data in turn.

\subsection{Mock Galaxy Catalogs}

Our model begins with a cosmological N-body simulation from which we
extract positions and masses of host dark matter halos within a
cosmological volume.  We populate the host halos with subhalos using
the method of \citet{Zentner05}.  Our method for producing mock
catalogs has been discussed in detail in \citet{Berrier06}.  We
provide a brief overview here, and refer the reader to
\citet{Berrier06} for a more detailed discussion.

Our numerical simulation was run with the Adaptive Refinement Tree
N-body code \citep{Kravtsov97} in a standard $\Lambda$CDM cosmology
with $\Omega_{m}=0.3$, $\Omega_{\Lambda} = 0.7$, H$_0$ = 70 km
s$^{-1}$ Mpc$^{-1}$, and $\sigma_8 = 0.9$.  The simulation followed
the evolution of $512^3$ particles in comoving box of $120$ h$^{-1}$
Mpc on a side, implying a particle mass of $m_p \simeq 1.1 \times 10^9$ h
$^{-1} M_\odot$.  The simulation grid was refined down to a minimum 
cell size of $h_{\mathrm{peak}} \simeq 1.8$ h$^{-1}$ kpc on a side.  
The simulation was previously discussed in
\citet{Tasitsiomi04}, \citet{Zentner05}, \citet{Allgood05}, and
\citet{Wechsler06}.  Host halos in the simulation were identified
using a variant of the Bound Density Maxima algorithm
\citep[BDM,][]{klypin_etal:99,Kravtsov04a}.  We define a halo virial
radius as the radius of the sphere, centered on the density peak,
within which the mean density is $\Delta_v (z)$ times the mean density
of the universe, $\rho_m$.  The virial overdensity $\Delta_v(z)$, is
given by the spherical top-hat collapse approximation and we compute
it using the fitting function of \citet{bryan_norman:98}.  Host halos 
are identified as those halos whose centers 
do not lie within the virial radius of 
another halo.  

In order to determine the substructure properties in each host dark
matter halo, we model their mass accretion histories and track their
substructure content using an analytic technique that exploits 
numerous simplifying approximations and several scaling relations 
derived from direct simulation \citep{Zentner05}.  For each host halo of mass
$M$ at redshift $z$ in our simulation volume, we randomly
generate a mass accretion history using the method of
\citet{Somerville99}.  At each merger event, we assign an initial
orbital energy and angular momentum to the infalling object according
to the probability distributions for these quantities derived from 
cosmological $N$-body simulations \citep{Zentner05}.  
Each accreted system becomes a subhalo at the time it is
accreted.  It is assigned a mass and a corresponding maximum circular
velocity at this time, \Vmax $=$ \Vin\ .  We integrate the orbit of
the subhalo in the potential of the main halo from the time of
accretion until $z=0$.  We model both tidal mass loss and internal 
heating of subhalos as well as dynamical friction using a modified form of the
Chandrasekhar formula \citep{chandrasekhar43} suggested by 
\citet{Hashimoto03}.  As subhalos orbit
within their hosts, they lose mass and their maximum circular
velocities decrease as the profiles are heated by interactions.
We remove galaxy subhalos from our catalogs once their maximum circular
velocities drop below \Vmax $= 80$~\kms.  This rough criterion is used 
to mimic the dissolution of the observable galaxy as a result of these 
interactions.

This procedure produces a population of subhalos within each host dark
matter halo in the volume.  In fact, the procedure is statistical and
relies on realizations of the small-scale density field and the
orbital parameters for infalling structure.  As such, each host may be
assigned numerous different subhalo populations that differ in their
detail due to realizing these statistical distributions with finite
samples.  We include such variations by producing four mock catalogs
from four independently-realized subhalo populations for each host.

The next step is to map galaxies onto the host halo and subhalo populations 
in the model.  Each subhalo has an
associated \Vin\ circular velocity that acts as a proxy for its
luminosity and each host halo has a corresponding ``central galaxy''
with a \Vin = \Vmax.  We assign luminosities to halos by matching
volume number densities of galaxies between the simulation and the
data; thus, we assume that luminosity is a monotonic function of \Vin.
In \citet{Berrier06}, we compared the close pair fraction predicted by
this method to close pair counts in the UZC and found good agreement
if we associated galaxy luminosity in a one-to-one way with \Vin.  We
use this association throughout this paper.  In addition to the 
work of \citet{Berrier06} which shows the two-point correlation
function extending down to $\sim$50 h$^{-1}$ kpc, a similar model 
has also been shown to reproduce the larger scale galaxy two-point
correlation function as a function of luminosity, scale, and redshift
\citep{Conroy06}.

As discussed below, we focus on halos with \Vin\ $> 160.5$ \kms\ in
order to define a mock galaxy catalog with the same number density as the
M$_{\rm B,j} \leq -19$ population we consider from the 2dF survey.
The numerical model predicts the number of galaxies with \Vin\ $>
160.5$ \kms\ that occupy every host halo in the simulation, \Nhost.
As discussed below, most galaxies in the simulation reside in halos by
themselves, with \Nhost\ $=1$, but galaxies in highly-clustered
regions tend to reside in massive host halos with multiple galaxies,
\Nhost\ $\gtrsim 9$.  In what follows we analyze the full 
simulation volume; thus, all results are appropriate for comparison to a
volume-limited sample.  In addition, we use all four mock catalogs 
constructed from the four, independent realizations of halo substructure 
for each host unless otherwise stated.

\subsection{Pairs in the 2dF Survey}

We examine star formation in pairs selected from the public 2dFGRS
database \citep{Colless01}.  In contrast to previous studies of 2dFGRS
pairs \citep[e.g.,][]{SolAlonso06}, we construct a volume-limited
sample with galaxies to $M_{Bj}$ = -19, assuming $\Omega_{\Lambda}$ =
0.7, $\Omega_{m}$ = 0.3, and $H_0$ = 70 km s$^{-1}$ Mpc$^{-1}$.  The
redshift range is from 0.010 to 0.0875.  We restrict the study to two
simple rectangles in the sky, with coordinates $148.97^o$ $\leq$
$\alpha$ $\leq$ $209.42^o$, $-4.78^o$ $\leq$ $\delta$ $\leq$ $2.21^o$
and $29.68^o$ $\leq$ $\alpha$ $\leq$ $50.94^o$, $-34.18^o$ $\leq$
$\delta$ $\leq$ $-25.32^o$.  While these strips are relatively
complete, there are some omitted regions near bright stars.  We only
select pairs from the portion of the sample not within 700 h$^{-1}$ kpc
of the edges of the rectangles on the sky or within 1000 km s$^{-1}$
of our redshift limits in order to probe the full environments of our
targeted objects.

The full sample covers a solid angle of 0.317 steradians and a volume
of 1.79 x $10^6$ h$^{-3}$ Mpc$^3$.  Integrating the luminosity
function reveals an expected object density of 0.0107 $\pm$ 0.0008
galaxies h$^3$ Mpc$^{-3}$ to ${\rm M_{Bj}}$ = -19 \citep{Croton05};
this density corresponds to a cutoff halo circular speed $V_{\rm in}$
= 160.5 km s$^{-1}$ in the simulation, which we use consistently
throughout the paper.  The targeted sample excluding the edges
includes 22,601 galaxies, with 1,344 galaxies in close ($\lesssim 50$
h$^{-1}$ kpc) pairs.

The primary problem faced in our data analysis is survey
incompleteness.  Overall, the 2dFGRS is approximately $86\%$ complete
\citep{Cross04}.  In a single observation, the 2dFGRS cannot
distinguish objects closer than 25 arcseconds on the sky.  At the
minimum, median, and maximum redshift of galaxies in the
volume-limited sample we consider, this fiber collision separation
corresponds to 4.8, 23, and 28 h$^{-1}$ kpc, respectively.  However, repeated
measurements of the same field with different fiber configurations
allow redshift measurements for some close pairs \citep{Lambas03,
SolAlonso04}.  In dense clusters, where galaxy pairs are
preferentially found, there are often many more objects than fibers,
leading to a differential incompleteness with environment that has
been characterized by the 2dFGRS survey team \citep{Colless01}.  We
use their tools to probe the effects of incompleteness in
\S~\ref{sec:data}.

Although it is not a primary focus of this work, we also verify the
results of our study using the {\it Sloan Digital Sky Survey} DR4 
\citep{Adelman-McCarthy06} {\it NYU Value-Added Galaxy Catalog} 
data \citep{Blanton05b}.  
We construct volume-limited catalogs to M$_{r} =
-19+5\log{h}$ and M$_{r} = -20+5\log{h}$ and consider both galaxies
with measured and unmeasured redshifts.  We use star formation
indicators including the star formation rate per unit mass from
\citet{Brinchmann04}.  In general, the SDSS results are in
qualitative agreement with the 2dF results we focus on here, but the
use of $r$-band selection and of different star formation indicators
introduce quantitative differences.

\section{The Environments of Galaxy Pairs} \label{sec:environments}

To isolate triggered star formation from star formation that is not
triggered by interactions, we must compare a well-defined pair sample
to the appropriate control sample --- typical examples of the
immediate progenitors of the interaction.  What types of galaxies
appear in pairs?  To answer this question, we investigate the
environments of pairs selected in our simulated volume-limited
redshift survey.

For each simulated ``galaxy'' (halo or subhalo) above our cutoff
circular velocity in the model, we compute \DN, the projected distance
to the object's nearest neighbor with \DeltaV$\le 1000$~km~s$^{-1}$.
We also measure \Nhost, and the total number of simulated galaxies
that lie within the same host dark matter halo as the object.
Typically, close pairs are defined based on small projected
separations, \DeltaD\ $< 30-100$ h$^{-1}$ kpc, with 50 h$^{-1}$ kpc a
commonly used value \citep{Barton00, Lin04, Berrier06}, within 
\DeltaV$=1000$~km~s$^{-1}$ of its neighbor.  Because we have full
three-dimensional information from the simulation, we can explore the
true nature of the ``apparent'' pairs selected with this technique.

\begin{figure}[t]
\epsscale{1.05} \plotone{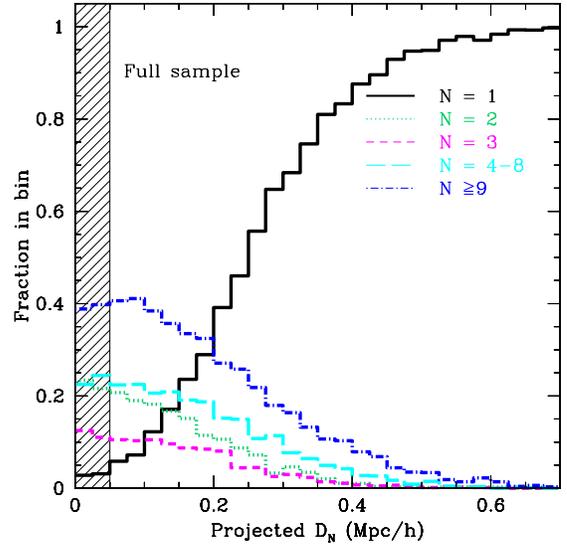}
\caption{The fraction of galaxies in the simulation with a total of
\Nhost\ galaxies in their host halos as a function of \DN, the projected
distance to its nearest neighbor.  For each galaxy in the mock
catalog, we measure \DN\ within \DeltaV $= 1000$~km~s$^{-1}$, and the
total number of galaxies in its host halo, \Nhost.  The different
lines correspond to different bins of \Nhost, \Nhost\ $= 1$ exactly ({\it
solid black}), 2 ({\it dotted green}), 3 ({\it dashed magenta}), 4-8
({\it long-dashed cyan}), and $\geq 9$ ({\it dot-dashed blue})
galaxies in the host halo.  We shade the ``pair zone,'' galaxies with
neighbors within 50 h$^{-1}$ kpc.  Close pairs are preferentially
found in very well-populated halos.}
\label{fig:Nsat_frac}
\end{figure}

We begin our exploration of close pair selection in
Figure~\ref{fig:Nsat_frac}.  To construct this figure, we first
compute \DN\ and \Nhost\ for each simulated galaxy, using the periodic
boundary conditions, if necessary, to fully sample its environment.
We then split the objects in each bin of \DN\ into subsets based on
the multiplicities of the halos in which the objects reside,
$N_{gal}(\mathrm{N}|\mathrm{D_N})$.  The ranges of the \Nhost\
bins are shown in the labels of Fig.~\ref{fig:Nsat_frac}.  Finally, at
each value of projected nearest-neighbor separation, we computed the
fraction of galaxies within that bin that are in the different subsets
of \Nhost.  Fig.~\ref{fig:Nsat_frac} shows this fraction among five
different bins of \Nhost\ as a function of \DN.  In each \DN\ bin, the
sum of all fractions is 1.  For reference, the shaded region
indicates the 50 h$^{-1}$ kpc ``pair zone.''  We note that the x-axis,
\DN, is the {\it directly} measurable quantity from redshift survey
data, while \Nhost\ is not directly measurable.

Fig.~\ref{fig:Nsat_frac} immediately reveals a major issue in
selecting appropriate control samples for galaxies in pairs.  {\it
Pairs in the simulation reside in a highly skewed distribution of
environments.}  In particular, pairs are vastly over-weighted toward
galaxies that reside within well-populated host dark matter halos
compared to the average.  For example, galaxies with \Nhost\ $\geq 9$
comprise 39\% of the pair sample but only 19\% of the simulation as a
whole.  In contrast, while isolated (N=1) galaxies make up 56\% of the
simulation as a whole, only 3\% of the apparent close pairs are
actually isolated galaxies.  These isolated galaxies in apparent pairs
are interlopers according to the analysis of \citet{Berrier06}.  The
abundant pairs in cluster and large-group systems are not necessarily
imminent mergers, nor are they necessarily even interacting directly
with each other.  \citet{Bailin07} note a similar problem in the
study of galactic satellite populations. 

From a qualitative perspective, Figure~\ref{fig:Nsat_frac} explains
many past observations regarding the star-forming properties of
galaxies in pairs.  For example, pair samples previously analyzed in
the 2dFGRS show an increased star formation rate relative to control
samples at close separations (\DeltaD $\lesssim 30$ h$^{-1}$ kpc)
where the interaction has a dominant effect \citep{Lambas03}.
However, at larger separations where the interaction is less important
($100 \lesssim {\rm \Delta D} \leq 200$ h$^{-1}$ kpc), the typical
pairs exhibit {\it less} star formation than the field
\citep[e.g.,][]{Lambas03, SolAlonso04, SolAlonso06}.  In our model,
these more widely-separated pairs are also in denser environments than
the field: 36\% are in \Nhost\ $\geq 9$-galaxy halos while only 19\%
are isolated.  This depressed star formation in widely-separated pairs
relative to the field results from the fact that galaxies living in
more massive and populated systems have suppressed star formation on
average.

\begin{figure}
\plotone{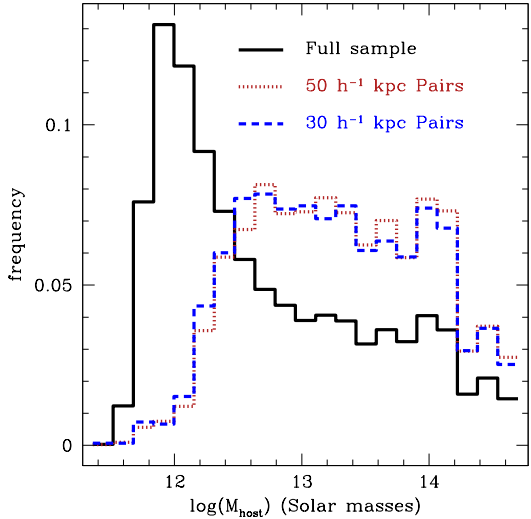}
\caption{The normalized distribution, ${\rm d} n/{\rm d}\log(M)$, of host halo
masses for 50 h$^{-1}$ kpc pairs ({\it dotted red}), 
closer 30 h$^{-1}$ kpc pairs ({\it dashed blue}) and the full sample ({\it solid
black}) in the simulation.  Pairs reside in much more massive halos
than typical ``field'' galaxies.}
\label{fig:Mhost_histogram}
\end{figure}

Because pairs are preferentially found in denser environments, the
naive comparison of star formation rates between pairs and {\it
typical} field galaxies will artificially underestimate any elevation
in star formation rate that is directly triggered by an interaction.
Fig.~\ref{fig:Mhost_histogram} shows the distribution of host halo
masses for galaxies in close (\DeltaD $\leq 50$ and 30 h$^{-1}$ kpc)
pairs and the sample as a whole.  This Figure emphasizes how skewed
the environments actually are.  The average host halo mass for the
full sample is ${\rm <M_{\rm host}> = 3.2 \times 10^{13}}$~h$^{-1}\
M_{\sun}$; for the close (50 h$^{-1}$ kpc) and closer (30 h$^{-1}$
kpc) samples of pairs it is 6.0 and 5.7$ \times 10^{13}\ {\rm h^{-1}\
M}_{\sun}$, respectively.  Moreover, the mode of the full sample,
$\sim 10^{12} \ {\rm h^{-1}\ M}_{\sun}$, is almost completely depleted
in the pair sample.  The analysis of the SDSS by \citet{Weinmann06}
tracks the dependence of galaxy properties on host halo mass.  We can
use their results to understand the implications for galaxies in
pairs, which are preferentially selected from higher-mass halos.
According to their results the ``average'' color of a galaxy in an
``average'' halo hosting a pair would be $\sim$0.05 magnitudes redder
in ${\rm g-r}$ color, have a specific star formation rate
\citep[e.g.,][]{Brinchmann04} that is $\sim$20\% lower, and have a
late-type fraction that is $\sim$5\% lower than the corresponding
averages for the field.  Thus, the suppressed star formation observed
in some widely-separated galaxy pairs is not related to a single
interaction, but is a selection effect associated with cluster and
group processes that suppress star formation.  This selection effect
must be accounted for in a fair analysis of triggered star formation.

\begin{figure}
\plotone{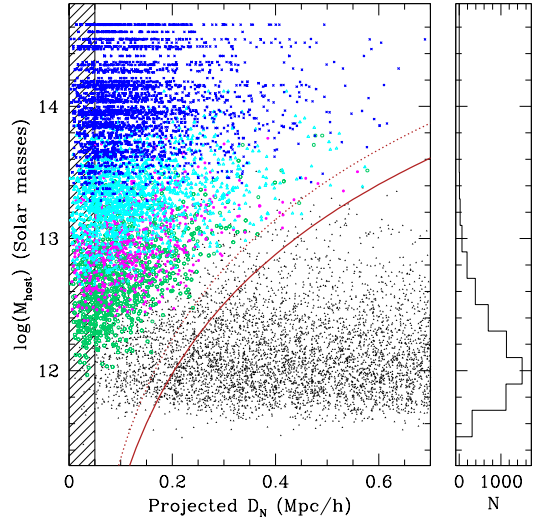}
\caption{The distribution of host halo mass as a function of distance
to the nearest neighbor, \DN, in the simulation for a single
realization of the substructure model.  ({\it Left}) For galaxies with
at least one luminous neighbor within 700 h$^{-1}$ kpc and \DeltaV $=
1000$~km~s$^{-1}$, we plot the log of the host halo mass as a function
of {\it projected} distance to the nearest neighbor in the simulation.
The point colors and types segregate galaxies based on halo
occupation.  We plot single-galaxy halos ({\it solid black
triangles}), \Nhost\ $= 2$ ({\it open green circles}), 3 ({\it filled
magenta circles}), 4-8 ({\it open cyan triangles}), and $\geq 9$-galaxy
halos ({\it blue crosses}).  The solid red line is the virial
radius of halos; the dotted red line is the virial radius reduced by a
factor of $\sqrt{2/3}$ to account for the typical difference between
projected and actual separation.  ({\it Right}) We also plot the host
mass histogram of the host halo of simulated galaxies with no
neighbors within 700 h$^{-1}$ kpc.}
\label{fig:Delta_D_vs_Mhost}
\end{figure}

Fig.~\ref{fig:Delta_D_vs_Mhost} illustrates the nature of the
differences between halos that host close pairs and ``typical'' dark
matter halos.  This figure plots galaxy host halo mass as a function
of projected separation to its nearest neighbor up to 700 h$^{-1}$ kpc
and within 1000 km s$^{-1}$ in our mock catalogs. The colors and point
types segregate host halos based on \Nhost, the number of galaxies
they host.  The structure of this plot provides a very general means
of viewing nearest-neighbor statistics.  The isolated halos line up
along the lower part of the plot; their locus is analogous to the
``two-halo'' term in the correlation function insofar as the pairs are
projections of two coincidentally aligned host halos.  The more
massive and populated dark matter halos line up along the left
vertical edge of the plot.  Again, pairs are an extremely skewed
population.  In contrast, galaxies that are isolated within
$\sim$400-500 h$^{-1}$ kpc are almost purely the only occupants in
their lower-mass halos.

In summary, a comparison of the properties of {\it typical} pairs and
typical isolated galaxies selected from the universe as a whole is not
the appropriate comparison for isolating star formation triggered by
galaxy interactions.  The ``field'' is dominated by isolated, low-mass
host halos with a mix of a few more luminous systems.  Pairs are
skewed, preferentially residing in higher-mass host halos.  Thus,
comparisons of the star-forming properties of pairs and the ``field''
do not isolate the effects of the recent interaction.  The properties
of close pairs reflect all of the other processes that occur in dense
group and cluster environments that may act to suppress star formation
during late-time interactions.  As such, naive comparisons will
dramatically underestimate the rise in star formation rate triggered
by galaxy interactions.  In the next section, we describe one means of
constructing samples that better isolate the effects of galaxy
interactions from other environment-related processes.

\section{Isolating triggered star formation} \label{sec:isolated}

Many different measures of large-scale environment have been 
developed to understand the properties of galaxies as a function of 
their surroundings.  With the advent 
of redshift surveys, these studies have focused on group-finding 
algorithms \citep{Huchra82}, {\it n}th nearest neighbor statistics 
\citep{Dressler80}, and local galaxy count measures
\citep[e.g.,][]{Blanton05a, Berrier07}.  Here, we adopt \Nseven, the
number of neighboring galaxies within 700 h$^{-1}$ kpc on the sky and
\DeltaV $= 1000$ km~s$^{-1}$ in redshift of the galaxy in question
\citep[see][]{Berrier07}.  We note that our results are qualitatively
insensitive to choices of environment statistic scale in the range of
700-1000 h$^{-1}$ kpc, but 700 h$^{-1}$ kpc yields a large enough
sample of galaxies in pairs in the 2dFGRS (\S~\ref{sec:data}). The use
of other statistics such as group-finding algorithms would yield
somewhat different results, but the basic problem of separating dense
systems from nearby but unassociated isolated dark matter halos
remains.

\begin{figure}[b]
\plotone{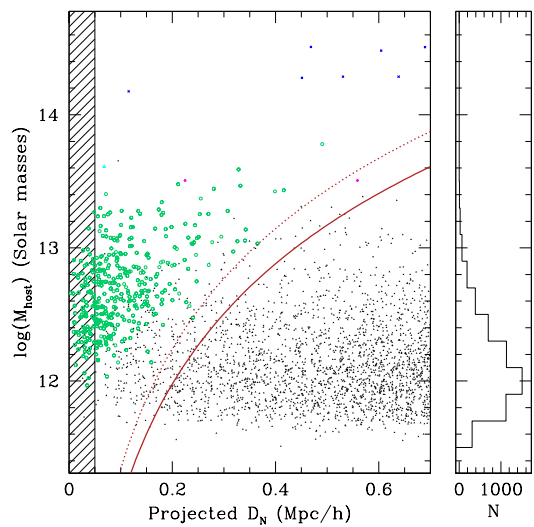}
\caption{The distribution of host halo mass as a function of distance
to the nearest neighbor, \DN, in the lowest-density environments for the 
objects in one realization of our mock galaxy catalogs.
The points are the same as Fig.~\ref{fig:Delta_D_vs_Mhost}, but restricted
to galaxies with exactly one neighbor within 700 h$^{-1}$ kpc and
\DeltaV $= 1000$ km~s$^{-1}$.  At right, we plot the host
mass histogram of galaxies with \Nseven\ $=0$, or a nearest neighbor
distance \DN\ $\geq 700$ h$^{-1}$ kpc.}
\label{fig:Delta_D_vs_Mhost_NMpc_eq_1}
\end{figure}

\begin{figure*}[t]
\plottwo{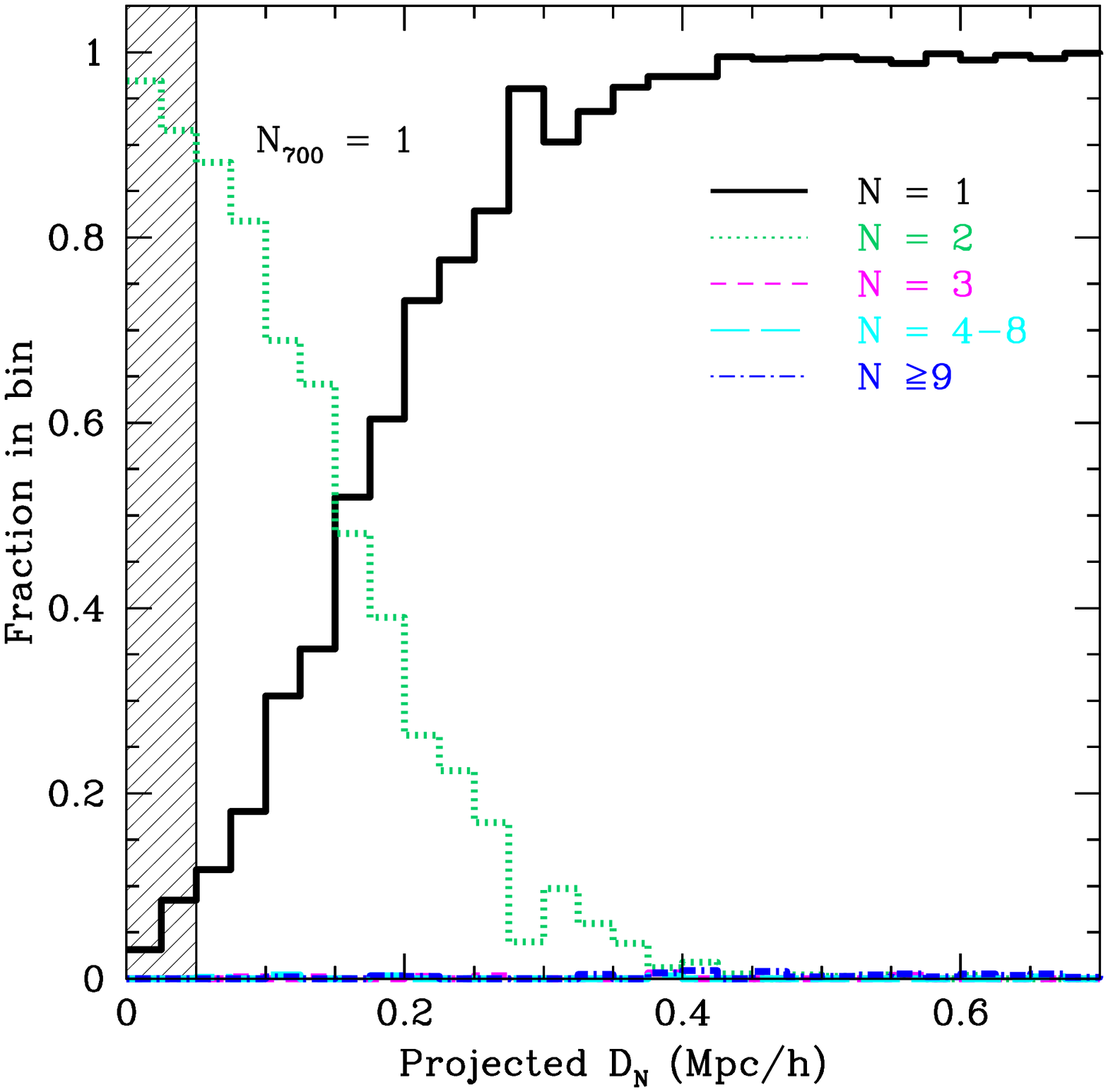}{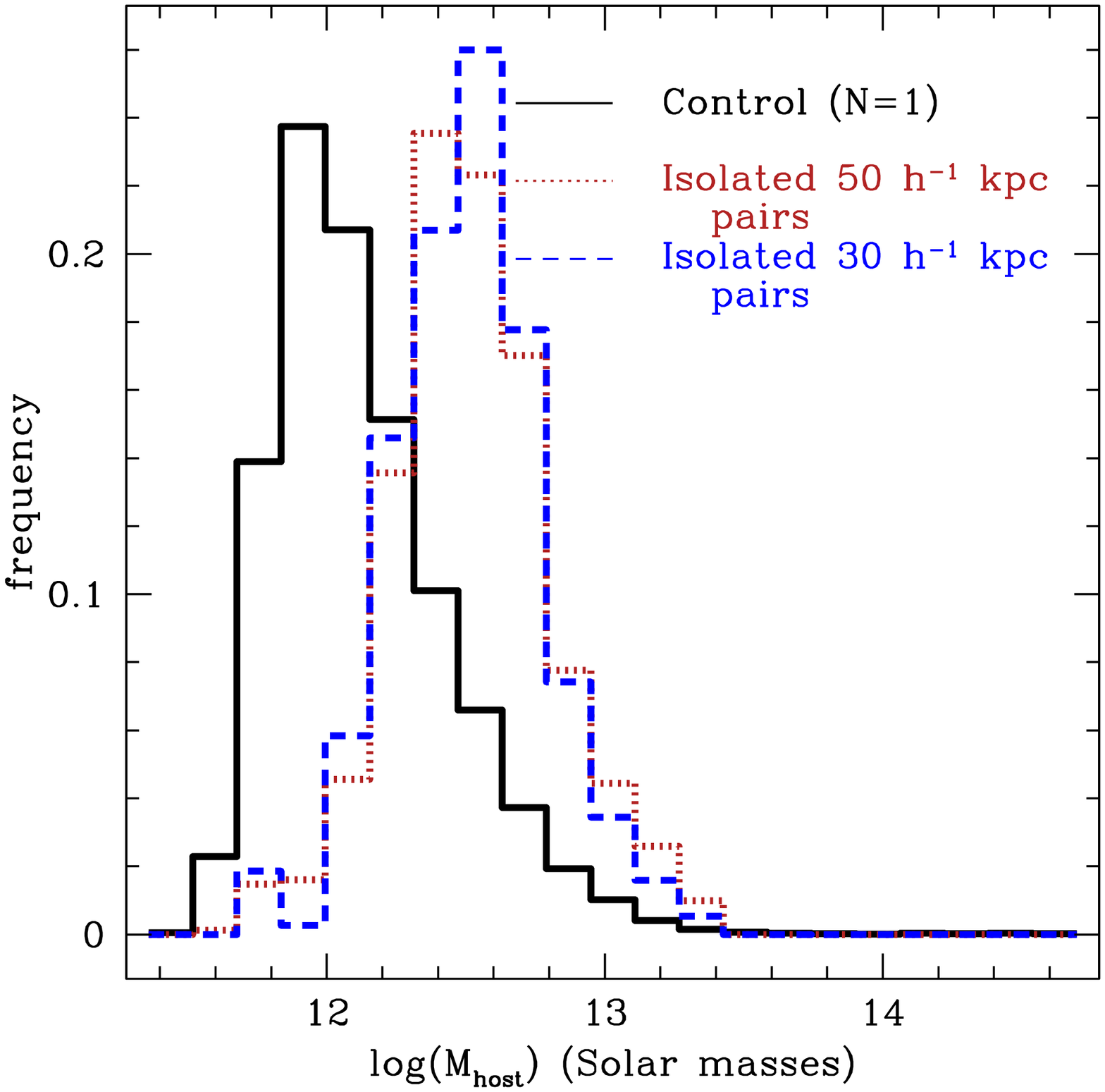}
\caption{The distribution of host halo masses and separations in
low-density environments in our mock catalogs.  We repeat
Figs.~\ref{fig:Nsat_frac} and \ref{fig:Mhost_histogram} restricting to
galaxies with \Nseven\ $= 1$, or exactly one luminous neighbor within
700 h$^{-1}$ kpc. In the left panel, we plot the fraction of galaxies
in hosts with a given number of members, \Nhost, as a function of the
distance to the nearest neighbor within 1000 km s$^{-1}$.  At right,
we plot histograms, ${\rm d} n/{\rm d}\log(M)$, of host halo masses
for the control sample of \Nseven\ $\leq 1$ galaxies with \DN\ $\geq
300$ h$^{-1}$ kpc ({\it black, solid}) and for $\leq 50$ h$^{-1}$ kpc
pairs with no other neighbors within 700 h$^{-1}$ kpc ({\it red,
dotted}) and similarly isolated $\leq 30$ h$^{-1}$ kpc pairs ({\it
blue dashed}).  The isolated pairs are twice as massive, on average,
as the isolated control sample, implying that the control galaxies are
likely the immediate progenitors of the pairs.}
\label{fig:frac_and_histogram}
\end{figure*}

Choosing galaxies with limited numbers of companions on $\sim$1~Mpc
scale environments is an effective way to preferentially select
galaxies in different types of dark matter halos
\citep[see][]{Focardi06}.  Fig.~\ref{fig:Delta_D_vs_Mhost_NMpc_eq_1}
shows the distribution of projected \DN\ as a function of host halo
mass for the the restricted set of environments corresponding to
galaxies with \Nseven\ $=1$.  These are galaxies that have {\it
exactly} one companion within 700 h$^{-1}$ kpc and 1000
km~s$^{-1}$.  This restriction is extremely clean and effective.
Widely-separated pairs are almost exclusively isolated halos and the
close pairs are almost purely \Nhost\ $=2$ systems.
Fig.~\ref{fig:frac_and_histogram} shows the fraction of host halos of
each type as a function of separation in this restricted sample and
the corresponding host halo masses in the pairs and full
distributions.  The fraction of \Nhalo\ $=2$ hosts goes from over 90\%
at the closest separations to a $50/50$ split of \Nhalo\ $=1$ and
\Nhalo\ $=2$ halos at $\sim$175 h$^{-1}$~kpc, to nearly 100\% \Nhalo\
$=1$ halos beyond 400 h$^{-1}$~kpc.

The key to the use of ultra-low-density environments is that the
restriction provides an extremely clean sample of pairs in \Nhost\
$=2$ halos at small separations and {\it a corresponding control
sample of isolated galaxies}.  This allows us to remove any effects
that may be due to numerous and repeated interactions in dense cluster
environments and study the relative effect of individual interactions.
The sample average masses are ${\rm <M_{\rm host}> = 4.2}$ and ${3.9 \times
10^{12}}$~h$^{-1}\ M_{\sun}$, respectively, 
for the objects that could be identified as isolated 50 h$^{-1}$ kpc
and 30 h$^{-1}$ kpc close pairs, and ${\rm <M_{\rm host}> = 2.2
\times 10^{12}}$~h$^{-1}\ M_{\sun}$ for the simulated isolated galaxy
sample with \DN\ $> 300$~h$^{-1}$~kpc and \Nseven\ $= 0$ or 1.  The
peaks in the distributions are offset by $\sim$0.5 dex.  Thus,
isolated close pair host halos are $\sim$2~--~3 times the typical
masses of the isolated galaxy (\Nhost\ $=1$) sample,
suggesting that the isolated galaxies are the appropriate immediate
progenitors of the \Nhalo\ $=2$ isolated close pairs.  In the
simulation, the subhalos we would identify observationally as isolated
close pairs were accreted roughly 0-8 Gyr ago, with an
average of 3.0 Gyr and a wide spread of $\sigma = 1.5$~Gyr.  For
comparison, the dynamical timescales for these halos are of order
$3$~Gyr.

%
%

As a guide to interpreting this isolated sample, consider a simple
example.  At present, the Milky Way and the Andromeda galaxy, viewed
from most projections, would be black points on the right side of
Fig.~\ref{fig:Delta_D_vs_Mhost_NMpc_eq_1}.  They are both isolated
with respect to {\it luminous} galaxies.  However, as they
move toward one another over the course of the next $\sim$few Gyr
\citep{Cox07}, they will move to the left on the plot.
Eventually their dark matter halos merge and  they 
become an \Nhost\ $=2$ halo, at which time they will move diagonally toward
the upper left and ``turn green.''  After perhaps 1-3 Gyr, the
baryonic galaxies will completely coalesce; the system would then move
horizontally to the right and ``turn black'' in the plot.  Isolated
systems like the Milky Way and Andromeda are the simplest units of
interaction and the best laboratories to isolate the effects of
interactions from other physical processes.  In the next section, we
find isolated pairs in the 2dFGRS and use the appropriate control and
paired samples to isolate the effects of an interaction.  As we show
in \S~\ref{sec:complex}, the situation is much more complex in denser
environments.

\section{Measuring triggered star formation} \label{sec:data}

\begin{figure}
\plotone{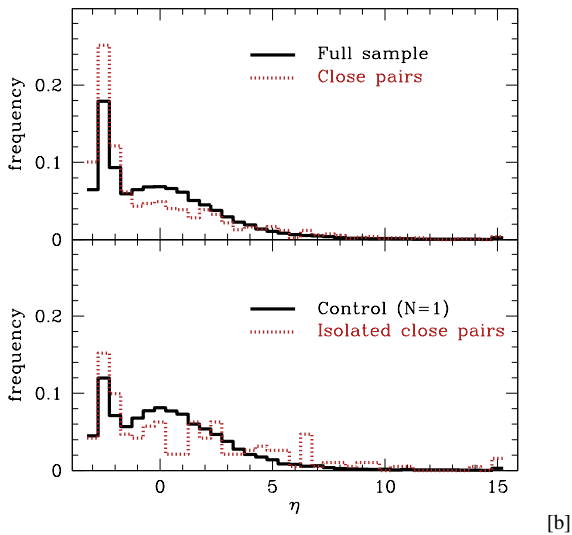}[b]
\caption{Star-forming properties of 2dFGRS galaxies in pairs.  ({\it
Top}) We plot the distribution of $\eta$ for the volume-limited sample
as a whole ({\it solid, black}) and for the close pairs sample ({\it
dotted, red}) with \DN\ $\leq 50$ h$^{-1}$ kpc.  Larger values of
$\eta$ correspond to higher star formation rates \citep{Madgwick03}.
Relative to the sample as a whole, the pairs exhibit {\it less} star
formation on average, as explained in Fig.~\ref{fig:Nsat_frac} by the
fact that they preferentially sample dense galactic
environments. ({\it Bottom}) We also restrict to isolated close pairs
({\it dotted, red}) and a control sample of galaxies with a \Nseven\
$\leq 1$ and \DN\ $\geq 300$ h$^{-1}$ kpc distant on the sky.  The
model shows that the pairs are almost always in \Nhost\ $=2$ halos
(93\%) and that 99.5\% of the control sample are in \Nhost\ $=1$
halos. Relative to the isolated ``control'' galaxies, the isolated
pairs have more galaxies on both ends of the $\eta$ distribution.}
\label{fig:eta}
\end{figure}

In \S~4, we demonstrate that the amount of triggered star formation
occurring in the universe cannot be measured by comparing typical
paired galaxies to typical field galaxies.  Dense galactic
environments are overrepresented in galaxies in pairs.  However, the
effects of an interaction can be identified independent from other
environmental processes by examining {\it isolated} pairs and
comparing them to {\it isolated} galaxies.  Although isolated pairs
are not the only environments in which triggered star formation is
important, they are the simplest to identify and study.

Here, we use 2dF survey pairs to examine
triggered star formation in galaxy pairs.  The ``full sample'' is the
full volume-limited sample that is far enough away from the edge of
the survey to sample the environments of galaxies without bias.  The
``close pairs'' sample is the subset of this full sample that has at
least one companion within 50 h$^{-1}$ kpc and 1000 km s$^{-1}$.  We
focus on the clean measure of the effects of interactions described in
\S~\ref{sec:isolated} by constructing the ``isolated close pairs''
sample of galaxies that have exactly one companion within 50 h$^{-1}$
kpc and 1000 km s$^{-1}$ and no others within 700 h$^{-1}$ kpc and
1000 km s$^{-1}$.  We also construct the ``control'' sample where
\Nseven\ $=0$ or 1 and the nearest neighbor is \DN\ $\geq 300$
h$^{-1}$ kpc away.

In the volume-limited sample of 41,239 galaxies, the ``full sample''
includes 22,601 galaxies that are not on the edges of the survey
volume.  There are 1344 galaxies in the ``close pairs'' sample, with
spectroscopic companions within 50 h$^{-1}$ kpc and 1000 km s$^{-1}$;
191 of these paired galaxies are in isolated close ($\leq 50$ h$^{-1}$
kpc) pairs and 72 are in isolated closer ($\leq 30$ h$^{-1}$ kpc) pairs.  
The isolated control sample with \DN\ $\geq 300$~h$^{-1}$
kpc and \Nseven\ $=0$ or 1 includes 8564 galaxies.

We examine the star-forming properties of galaxies using the spectral
parameter $\eta$, which \citet{Madgwick02} derive from principal
component analysis (PCA) of the 2dFGRS as a whole. Their PCA analysis
shows that two-thirds of the variance in the 2dFGRS spectra in a
volume-limited sample are contained in the first two projections.
$\eta$ is directly measured from a combination of these two
projections and is readily available in the 2dFGRS database.
\citet{Madgwick03} show that $\eta$ is closely related to both the
H$\alpha$ equivalent width measured from spectra and to the stellar
birthrate parameter, $b$, the present star formation rate of the
galaxy divided by its average past star formation rate.  Higher values
of $\eta$ correspond directly to higher current star formation rates.
\citet{Madgwick03} use model spectral energy distributions with a wide
variety of star formation histories to explore this correlation
between $\eta$ and $b$.

In the top of Fig.~\ref{fig:eta} we plot the distribution of $\eta$
for the sample as a whole and all the close pairs. The pairs as a
whole exhibit, on average, {\it less} overall star formation than the
full volume-limited galaxy sample.  The average for the full sample
and the close pairs sample are $<\eta>=0.07$ and -0.22,
respectively. Fig.~\ref{fig:Nsat_frac} demonstrates why.  In 
the model, most (56\%) galaxies are alone in their dark matter
halos.  In $\leq 50$ h$^{-1}$ kpc pairs, however, nearly all (97\%)
of the galaxies are in denser systems, and often much denser
systems. Because pairs preferentially reside in extremely dense
environments, their star formation is suppressed except when they are
strongly interacting.

As we demonstrate in \S~\ref{sec:isolated}, the model predicts that
99.5\% of the galaxies in the isolated control sample are alone in their
dark matter halos and that 93\% of the isolated pairs are in \Nhalo\
$=2$ halos.  
Thus, the bottom of Fig.~\ref{fig:eta} provides us with a
nearly direct measure of the effects of an interaction, by allowing us
to compare the true progenitors (the \Nhalo\ $=1$ ``control'') of the
paired (\Nhalo\ $=2$) systems.  A Kolmogorov-Smirnov test indicates
that the control and close-pair $\eta$ distributions differ, with
P$_{\rm K-S} = 6.1 \times 10^{-4}$.





\begin{figure}
\plotone{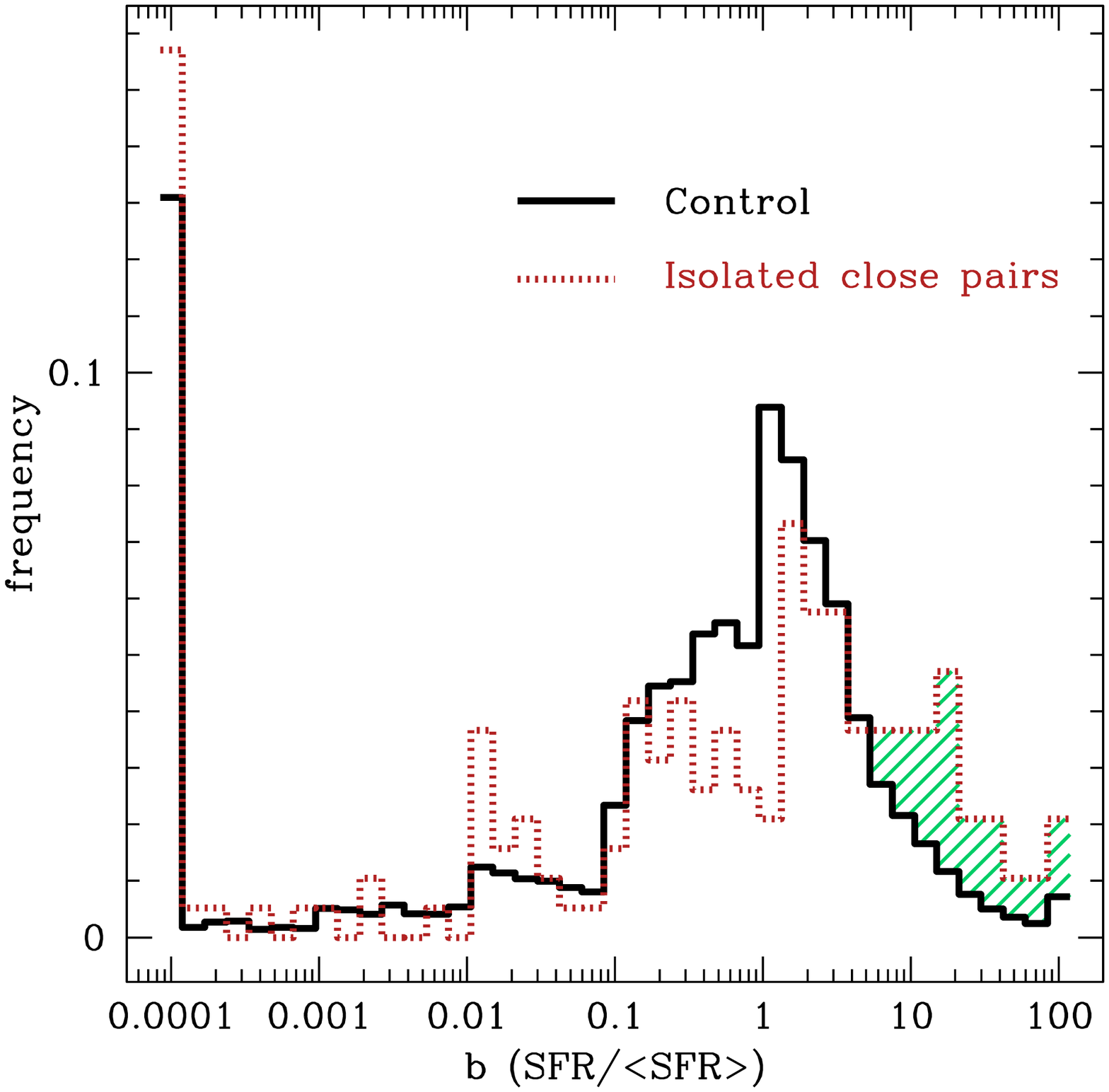}[b]
\caption{Normalized star formation rates of 2dFGRS galaxies in pairs.
We plot the distribution of $b$, the ratio of the current star
formation rate to the average past star formation rate, inferred from
$\eta$ \citep[see][]{Madgwick03}.  We restrict to isolated close pairs
({\it dotted, red}) and a control sample of galaxies with \Nseven\
$\leq 1$ and \DN\ $\geq 300$ h$^{-1}$ kpc.  A Kolmogorov-Smirnov test
indicates that the control and close-pair $\eta$ distributions differ,
with P$_{\rm K-S} = 6.1 \times 10^{-4}$.  Relative to the \Nhalo\ $=1$
``control'' galaxies, the isolated pairs have fewer galaxies forming
stars at their historical average rate; pairs exhibit an excess of
galaxies that are very deficient in star formation (5\% excess with $b
\leq 0.1$) and an excess of galaxies ({\it green shaded region})
with rates boosted by factors of
$\gtrsim 5$ (14\%) where the high-$b$ excess, or the
shaded region, has an average $\left< b \right> \sim 30$.  }
\label{fig:b}
\end{figure}

\begin{figure}
\plotone{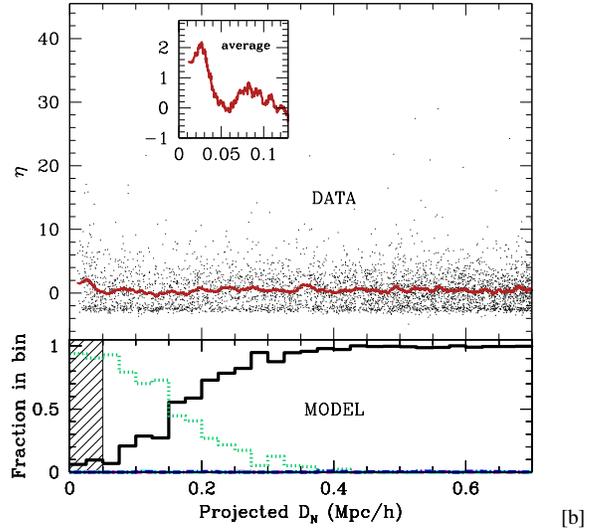}[b]
\caption{Star formation as a function of distance to nearest neighbor
in sparse environments.  ({\it Top}) For the 2dFGRS sample with
\Nseven\ $=1$, we plot $\eta$ as a function of separation to the
nearest neighbor; the red line is the averaged smoothed by 100 points.
The inset displays the average for \DN\ $\lesssim 0.13$ 
h$^{-1}$ Mpc, highlighting the signal of triggered star formation
at \DN\ $\lesssim 0.03$ h$^{-1}$ Mpc.
({\it Bottom}) For the \Nseven\ $=1$ sample in the simulation, we plot the
fraction of \Nhalo\ $=1$ ({\it solid, black}) and \Nhalo\ $=2$ 
({\it dotted, green}) galaxies as a function of \DN, as in 
Fig.~\ref{fig:frac_and_histogram}.}
\label{fig:dd_eta_n1}
\end{figure}

In Fig.~\ref{fig:b}, we use the approximate values in Figure 7 of
\citet{Madgwick03} to convert $\eta$ into $b$, the ratio of the
current star formation rate to the average past star formation rate,
for the isolated control and isolated pairs samples only.  Relative to
the \Nhalo\ $=1$ ``control'' galaxies, the isolated pairs have fewer
galaxies forming stars at their historical average rate ($b \sim 1$).
Pairs include an excess of low- and high-star-formation-rate galaxies.
While 24\% of the control sample are forming stars with $b < 0.1$,
29\% of the $< 50$ h$^{-1}$ kpc pairs and 26\% of the $< 30$ h$^{-1}$
kpc are forming stars at these low rates; thus, 50 h$^{-1}$ kpc (30
h$^{-1}$ kpc) pairs have a 5\% (2\%) excess of slow (or non-) star
formers.

At the high end of the star-formation rate distribution, 10\% of the
control (\Nhalo\ $=1$) sample have rates boosted by $\gtrsim$ a factor
of 5.  In contrast, 50 h$^{-1}$ kpc (30 h$^{-1}$ kpc) pairs are
boosted by $b \geq 5$ 24\% (30.5\%) of the time, for a 14\% (20\%)
excess of high star formation rate galaxies.  Binning the data with
fine bins, weighting by $b$, and limiting the largest (extrapolated)
boost to $b=100$, the high-$b$ ($b > 5$) excess in the pairs, the
shaded region in Fig.~\ref{fig:b}, has an average of $b \sim 31$ (34).
In other words, statistically, this technique shows that for the
population of pairs with triggered star formation, the average boost in
the star formation rate over the average past rate in that galaxy is a
factor of $\sim$30.

The error in the average boost due to triggered star formation is
dominated by the scatter in the relationship between the physical
parameter $b = {\rm SFR/\left< SFR \right>}$ and the measured $\eta$.
We estimate the error introduced by this scatter using a Monte-Carlo
simulation.  \citet{Madgwick03} explore the scatter in the theoretical
relationship between $b$ and $\eta$ using synthetic spectra with star
formation histories predicted by semi-analytic models.  For a typical
distribution of star formation histories, this scatter is substantial.
We adopt the 1-$\sigma$ scatter in the $\eta$-$b$ distribution in
\citet{Madgwick03} in our Monte Carlo simulation, drawing deviations
at random from the average $b$ for a given (measured) $\eta$.  We
resample the entire $b$ distribution for the control, pair, and very
close (30 h$^{-1}$ kpc) pair $\eta$ distributions and remeasure the
weighted average boost of the excess strong star formers with $b > 5$
in the pairs.  Because the scatter of $b$ as a function of $\eta$ is
much larger in the high-$b$ direction, the 1-$\sigma$ range of
resulting values of the average boost of triggered star formation is
42 -- 65 for the 50 h$^{-1}$ kpc pairs and 39 -- 61 30 h$^{-1}$ kpc
pairs.  We conclude from this analysis that the large scatter between
$b$ and $\eta$ results in a wide range of possible star formation
boosts resulting from triggered star formation.  This type of
uncertainty would be typical for other popular parameterizations of
star formation history derived from optical spectra, as well.  Many
optical spectral measures of star formation break down in the
``bursty'' star formation regime that is typical of interacting
galaxies.

The excess fraction of starbursting galaxies is enhanced in even
closer (30 h$^{-1}$ kpc) pairs, while the excess of low star formation
rate galaxies is not.  We explore the dependence of starbursting
galaxies on separation further by dividing the data into bins based on
\DN.  With 50 h$^{-1}$ kpc bins, only the \DN\ $\leq 50$ h$^{-1}$ kpc
bin has a statistically significant excess of high-$b$ galaxies
compared with the \Nhalo\ $=1$ control; with 25 h$^{-1}$ kpc bins,
only the \DN\ $\leq 50$ h$^{-1}$ bins show a significant excess.

\section{Understanding the trends} \label{sec:trends}

These trends with separation on the sky have been noted previously by
several authors using many different pair datasets \citep{Barton00,
Lambas03, Nikolic04, SolAlonso04, SolAlonso06, Lin07}.  Thus, the fact
that we reproduce them here is not surprising.  However, as
Figs.~\ref{fig:Nsat_frac} and \ref{fig:frac_and_histogram} show, all
pair samples suffer from systematic environmental differences as a
function of separation on the sky that are not directly related to
interactions.  Typical galaxy samples with no environmental controls
are dominated by cluster galaxies at separations $\lesssim 200$
h$^{-1}$ kpc and field galaxies at separations $\gtrsim 200$ h$^{-1}$
kpc.  Even the most isolated samples of pairs are dominated by pairs
in \Nhalo\ $=2$ halos at separations $\lesssim 200$ h$^{-1}$ kpc and
by \Nhalo\ $=1$ halos at separations $\gtrsim 200$ h$^{-1}$ kpc.
Here, for the first time, we are uniquely poised to explore these
effects and construct appropriate control samples using models.

Although we note that isolated galaxies in pairs are the simplest
interactions to study and compare with isolated galaxies, they still
may be affected by other processes.  \Nhalo\ $=2$ halos are often the
result of the merger of two \Nhalo\ $=1$ systems, but they can also
result from an \Nhalo\ $=3$ system after two of the three baryonic
galaxies coalesce.  An origin as a \Nhalo\ $=3$ system is, however,
much less likely for a given system.  Although our models
are not presently capable of tracking these mergers in detail,
we note that the density of \Nhalo\
$=3$ is only 8\% of the density of \Nhalo\ $=1$ systems.  Thus, the
simplest interpretation of the result in Fig.~\ref{fig:b} is that the
5\% excess of low-star-formation systems in the \Nhalo\ $=2$
galaxies results from the fact that galaxies in \Nhalo\ $=2$ halos are
more likely than isolated galaxies to have experienced previous
gas-consuming interactions, and that the 14\% excess of strongly
star-forming galaxies in \Nhalo\ $=2$ halos results {\it directly}
from recent interactions.  Future investigation of these models will
clarify this evolution.

\begin{figure}[t]
\plotone{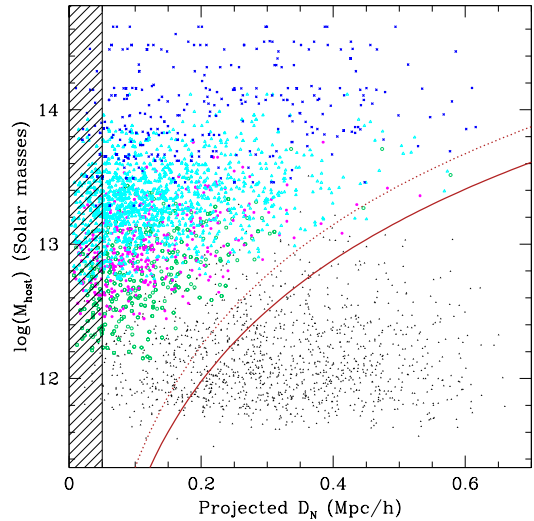}
\caption{Denser environments: the distribution of host halo 
mass as a function of distance to the nearest neighbor, \DN, but
restricted to galaxies with 3-8 neighbors within 700 h$^{-1}$ kpc and
\DeltaV $= 1000$ km~s$^{-1}$.  The points are the same as
Fig.~\ref{fig:Delta_D_vs_Mhost}.}
\label{fig:Delta_D_vs_Mhost_NMpc_38}
\end{figure}

Supporting this picture, the boosted star formation does not extend to
all systems with \Nhalo\ $=2$.  In Fig.~\ref{fig:dd_eta_n1}, we plot
$\eta$ as a function of separation for the isolated (\Nseven\ $=1$)
sample.   A high mean $\eta$ persists only out to 
$\sim$40-50 h$^{-1}$ kpc.  However, the model plotted below the
points for clarity shows that the \Nhalo\ $=2$ halos dominate well
past 100 h$^{-1}$ kpc.  Thus, the boosted star formation at small
separations results from the physical interaction and not merely the
different mix of halo types at smaller separations.  
A boosted {\it average} star formation rate does not persist
beyond a few tens of h$^{-1}$ kpc, probably because the strength of
the star formation fades as the burst ages \citep[see][]{Barton00,
Barton03}, and because the fraction of galaxies that have had a recent 
close pass is a strong function of \DN. 

The halo substructure model tracks close pericentric passages between
satellites and the central galaxy in the dark matter host halo.  These
simulations show that for the isolated close pair galaxies in true
\Nhalo\ $=2$ systems, 12\% of the $\leq$50 h$^{-1}$ kpc pairs and 16\%
of the $\leq$30 h$^{-1}$ kpc pairs have had a close pass with R$_{\rm
peri} < 30$~h$^{-1}$ kpc within the last Gyr.  In addition, 40\% of the 50
h$^{-1}$ kpc pairs and 45\% of the 30 h$^{-1}$ kpc pairs have had a
pass with R$_{\rm peri} < 50$~h$^{-1}$ kpc within the last 1.2 Gyr.
Thus, the 2dF data are consistent with a range of scenarios, including
at its extremes, (1) the possibility that nearly all galaxies with $<
30$~h$^{-1}$ kpc passes burst by an average factor of $\sim$30 for 1
Gyr, or (2) the possibility that $\lesssim$ half of the galaxies with
$< 50$~h$^{-1}$ kpc passes burst this much for $\sim$1.2 Gyr.  If star
formation continues for these long timescales, more widely-separated
pairs will contain some systems with triggered star formation.
However, the average star formation rate at wider separations is not
affected because pairs that have had close passes are diluted.  The
model shows that only 3\% of the true \Nhalo\ $=2$ systems with
separations $100 < $ \DN\ $ < 200$ h~$^{-1}$ kpc have had a close ($<
30$ h$^{-1}$kpc) pass within the last Gyr.

We will attempt further exploration of these scenarios in later work.
However, preliminary analysis of {\it Sloan Digital Sky Survey} DR4
{\it NYU Value-Added Galaxy Catalog} data \citet{York00, Blanton05b,
Adelman-McCarthy06} --- albeit with quite different star formation
measures --- suggests that the bursting fraction is a strong function
of the cutoff luminosity of the sample studied.  The fraction also
depends in detail on the star formation measure and on the luminosity
cutoff used for constructing the volume-limited sample, probably
because star formation rate is a strong function of galaxy
luminosity. Qualitatively, the results remain quite consistent between
the 2dF and SDSS samples.

The optical spectra available in the 2dFGRS (and the SDSS) do
not measure the amount of star formation that is embedded in dust
clouds.  Thus, the typical boost of the high-$b$ excess provides an
approximate lower limit to the amount of triggered star formation in a
close galaxy-galaxy pass.  As \citet{Cox06} demonstrate,
high-resolution hydrodynamical simulations predict a wide range of
star formation boosts from these close passes.  Their models predict
SFR boosts of 6~--~20 above a galaxy's isolated star formation rate in
prograde encounters during the $\sim$ Gyr between the first close pass
and the final merger.  Surprisingly, our estimate of $<b> \sim 30$
averaged over the galaxies that have triggered star formation exceeds this
range.  However, the uncertainties associated with the mapping between
$\eta$ and $b$ are substantial.  Thus, this 2dF analysis remains
broadly consistent with the models described in \citet{Cox06} and does
not yet distinguish among them.

\subsection{The effects of survey incompleteness}

The 2dFGRS is not spectroscopically complete; the survey is especially
deficient in close pairs.  We examine the effects of incompleteness
using the measured completeness of the sectors in which the galaxies
fall, and the 2dF parameters ``wsel,'' a weighted measure of whether
the ten nearest galaxies have high-quality redshifts, and ``bjlim,''
the limiting magnitude of the relevant sector \citep{Colless01,
Norberg02}.  Restricting the sample to sectors with known completeness
$> 90\%$ and with wsel $\geq 1.3$ and bjlim $\geq$19.35, we find a
reduced isolated pairs sample with only 67 pair galaxies and 2132
control galaxies.  As a result, a K-S test no longer reveals any
statistical significance of the difference in $\eta$ distributions
(P$_{\rm K-S}=0.35$).  Nevertheless, the differences in star-forming
properties of the isolated control and isolated remain qualitatively
unchanged: in the restricted control sample, 12\% of the galaxies have
$b > 5$; in the restricted isolated pairs, 21\% have $b > 5$.  Thus,
contamination of the isolated control sample appears to have no
significant effects on our results.

We further test the effects of redshift survey completeness by
applying the same analysis to the SDSS DR4 sample \citep{Blanton05b,
Adelman-McCarthy06}.  Although the detailed results are beyond the
scope of this paper, we are able to use SDSS to find a pure isolated
pair sample and a pure control sample with SDSS.  We construct these
samples because we are able to exclude all systems with neighbors that
have unmeasured redshifts. The measured excess of star formation in
isolated pairs compared with the control sample, from SDSS depends
sensitively on the cutoff luminosity of the volume-limited sample.
This dependence arises because higher-luminosity galaxies have less
star formation in general.  Using H$\alpha$ equivalent width as a
measure of star formation, the measured excess of pairs with
EW(H$\alpha$) $\geq 25$~\AA\ ranges from somewhat lower than the 2dF
results we present here (4\%) for a sample volume limited to M$_{r} =
-20+5\log{h}$ to 19\% for a sample volume limited to M$_{r} =
-19+5\log{h}$.  The 2dF luminosity limit corresponds to galaxies that
are midway between these numbers.  Thus, the 2dF and SDSS analyses
agree qualitatively, and the fraction of galaxies in close pairs with
triggered star formation increases when lower luminosity galaxies are
included.  The differences may result from the use of very different
measures of star formation or from the use of galaxies in an
intrinsically different luminosity range.  However, this analysis
suggests that ``impurity'' in the 2dF sample --- the inclusion of
galaxies and pairs that are not truly isolated --- is not biasing our
results significantly.

\section{More complex environments} \label{sec:complex}

Isolated pairs of galaxies are the easiest systems to construct an
appropriate control sample for.  However, isolated pairs are by no
means the only pairs that show evidence for triggered star
formation.  \citet{SolAlonso06} show that $b$ still increases as pair
separation on the sky decreases for close pairs in denser
environments.

In Sec.~\ref{sec:isolated}, we describe the effectiveness of isolating
the lowest-density environments to create an appropriate control
sample for galaxies in isolated pairs.  Unfortunately, the same type
of technique does not work for higher-density environments.  One 
cannot separate galaxies into a density bin and 
expect the non-pairs to be a good control sample for the pairs.

To illustrate this point, we restrict the sample to intermediate
values of our environment statistic; in
Fig.~\ref{fig:Delta_D_vs_Mhost_NMpc_38} we plot galaxies with $3 \leq
$~\Nseven~$\leq 8$.  As expected, isolated \Nhalo\ $=1$ halos are
missing from the 50 h$^{-1}$ kpc pairs sample (3.1\%) but still
constitute a significant fraction of the overall sample (31\%).
This result immediately illustrates the problem with breaking sets of
galaxies into low, intermediate, and high-density samples and
separating the pairs from these samples \citep[e.g.,][]{SolAlonso06}.
The close pairs have a completely different environmental mix from the
rest of this intermediate-density sample.  The mix is a strong
function of distance to the nearest neighbor, and even the pairs
reside in a huge range of hosts, from halos with \Nhalo\ $=2$ galaxies
to halos with \Nhalo\ $> 9$.  However, as \citet{SolAlonso06} show,
pairs in dense environments do exhibit a rise in star formation
rate with a very close encounter.

The ineffectiveness of comparing pairs and ``control'' galaxies in
intermediate- or high-density environments is not merely a result of
the exact technique we use.  Group-finding algorithms can be tuned to
lower the incompleteness and/or impurity rates on various scales
\citep[e.g.,][]{Yang05,Gerke05,Weinmann06,Berlind06,Koester07}, but
these algorithms will always include isolated galaxies in denser
systems from the control sample at a higher rate than they appear in
the pairs.  Because the distribution of star formation rates in
isolated galaxies has a high-star-formation tail, their contamination
will always lower the measured difference in star formation rate
between paired and non-paired galaxies in loose groups.  One effective
approach in these environments is to estimate the contamination of
\Nhalo\ $=1$ galaxies to the sample of interest and to statistically
subtract its contribution.

What is the appropriate control sample for pairs in dense
environments?  Even if one could identify the host halo mass and
occupation of galaxies perfectly, this would still be a difficult
question to answer.  The immediate progenitors of pairs in dense
environments can range from isolated galaxies that have just fallen into 
the system to galaxies that have been in a loose group for along time but
are just encountering one another for the first time.  Tracking the
evolution of the orbits of subhalos in the models will reveal the
progenitors of interactions in crowded environments, but that analysis
is beyond the scope of the present paper.

\section{Conclusion} \label{sec:conclusion}

Here, we use mock catalogs based on simulations of cosmological structure 
formation in the prevailing $\Lambda$CDM model 
to understand the properties of galaxies in pairs.
We examine the typical host dark matter halos of galaxies in
apparent pairs in the simulations and find that:

\begin{enumerate}

\item Our simulations show that galaxies in close pairs are preferentially located in cluster
and group environments.  As a result, typical close pairs are not
ideal for understanding triggered star formation in galaxy-galaxy
interactions.  This result explains why galaxies in pairs can appear
redder or can appear to have less star formation than the typical
``field'' galaxies.  If the goal is to isolate the effects of a recent
interaction, the ``field'' is not an appropriate control sample with
which to compare typical galaxies in close pairs.

\medskip

\item Using the simulations, we show that 
close pairs in the sparsest environments, with only one neighbor
within 700 h$^{-1}$ kpc, provide the simplest situation for isolating
the effects of an interaction.  Close ($< 50$ or $< 30$ 
h$^{-1}$ kpc) isolated
pairs are almost exclusively in \Nhalo\ $= 2$ halos (93\%).  Very
isolated galaxies with at most one neighbor within 700 h$^{-1}$ kpc
that is at least \DN\ $\geq 300$ h$^{-1}$ kpc away are almost
exclusively in \Nhalo\ $=1$ halos (99.5\%).  Thus, isolated field
galaxies are the ideal control sample of the progenitors of isolated
pairs.

\item We study isolated pairs and galaxies in a volume-limited sample
of M$_{\rm B,j} \leq -19$ galaxies from the 2dFGRS.  Isolated pairs
with \DeltaD\ $\leq 50$ h$^{-1}$ kpc ($\leq 30$ h$^{-1}$ kpc) show an
excess of both strongly star-forming galaxies and of non-star-forming
galaxies when compared to \Nhalo\ $=1$ control sample.  While 24\% of
the control sample is forming stars at $b \lesssim 0.1$ of its average
past rate, 29\% (26\%) of the close pairs lack star formation to this
extent.  In addition, 24\% (30.5\%) of the pairs are forming stars at
$b \gtrsim 5$ times their average past rate while only 10\% of the
control sample has rates this high.  The rapidly-star-forming excess
population in the pairs is almost certainly due to the direct effects
of the interaction.  However, we note that the fraction of galaxies
undergoing triggered starbursts is a function of the limiting absolute
magnitude of the sample because a much smaller fraction of more
luminous galaxies are able to form stars at all.

\item For the isolated close galaxy pairs in the 2dFGRS, 
the galaxies with triggered
star formation have an average star formation boost of $b = $
SFR/$\left<{\rm SFR}\right>\sim 30$.

\item The measurement of the optical boost of $b\sim30$ for
interactions in 2dFGRS is approximate, due in large part to the large
scatter in relationship between star formation history and measured
optical spectral parameters.  The measure also excludes embedded star
formation that is not detectable at optical wavelengths.  Thus, it is an
estimate of the lower limit to the boost in star formation rate caused
by triggered star formation.  This boost is higher than predictions
derived for boosts from triggered star formation using hydrodynamic
simulations \citep[e.g.,][]{Cox06}, although the uncertainties are too
large to rule out specific models at this stage.

\item By tracking orbits in the substructure model, we show that for
the isolated close pair galaxies in true \Nhalo\ $=2$ systems, only
12\% (16 \%) of the 50 (30) h$^{-1}$ kpc pairs have had a close pass
with R$_{\rm peri} < 30$~h$^{-1}$ kpc within the last Gyr, and 40\%
(45\%) have had a pass with R$_{\rm peri} < 50$~h$^{-1}$ kpc within
the last 1.2 Gyr.  Thus, the detection of triggered star formation in
14\% (20\%) of these L$^{\star}$ and sub-L$^{\star}$ systems in the 2dF data
suggest that a large fraction of the galaxies that experience close
passes respond with triggered star formation.

\end{enumerate}

\acknowledgements We gratefully acknowledge the hard work of the
2dFGRS team and thank them for making their data public.  We also
gratefully acknowledge use of the 2dFGRS mask software by Peder
Norberg and Shaun Cole.  We thank Joel Berrier and Heather Guenther
for helping to lay the foundations for the use of these cosmological
models to study galaxies in pairs.  We thank Margaret Geller, Alison
Coil, Joel Primack, Sara Ellison, and an anonymous referee for giving
us many insightful and useful comments at various stages. We thank
Anatoly Klypin for running the numerical simulation used here, which
was performed on the IBM RS/6000 SP3 system at the National Energy
Research Scientific Computing Center (NERSC).  EJB, JAA, and JSB
acknowledge support from the Center for Cosmology at UC Irvine; JSB is
supported by National Science Foundation (NSF) grant AST-0507916.  ARZ
is supported by the NSF Astronomy and Astrophysics Postdoctoral
Fellowship Program under grant AST-0602122, by the Kavli Institute for
Cosmological Physics at The University of Chicago, and by NSF PHY
0114422.  RHW is supported in part by the U.S. Department of Energy
under contract number DE-AC02-76SF00515.

\bibliography{ms.bbl}

\end{document}